\documentclass[aps,superscriptaddress,showkeys]{revtex4}
\usepackage{amsfonts,amssymb,amsmath,latexsym,epsfig} 
\begin{document}

\title{Transport Properties of Random Walks\\ on Scale-Free/Regular-Lattice Hybrid Networks}

\author{Juli\'an Candia}
\affiliation{Consortium of the Americas for Interdisciplinary Science and Department of\\
Physics and Astronomy, University of New Mexico, Albuquerque, NM 87131, USA}
\affiliation{Center for Complex Network Research and Department of Physics,\\ 
University of Notre Dame, Notre Dame, IN 46556, USA}
\author{Paul E. Parris}
\affiliation{Department of Physics, 
University of Missouri-Rolla, Rolla, MO 65409, USA}
\author{V.M. Kenkre}
\affiliation{Consortium of the Americas for Interdisciplinary Science and Department of\\
Physics and Astronomy, University of New Mexico, Albuquerque, NM 87131, USA}

\begin{abstract}
We study numerically the mean access times for random walks on hybrid
disordered structures formed by embedding scale-free networks into
regular
lattices, considering different transition rates for steps across
lattice
bonds ($F$) and across network shortcuts ($f$). For fast
shortcuts ($f/F\gg 1 $) and low shortcut densities, traversal
time data collapse onto an
universal curve, while a crossover behavior that can be related to the
percolation threshold of the scale-free network component is identified
at
higher shortcut densities, in analogy to similar observations reported
recently in Newman-Watts small-world networks. Furthermore, we observe
that
random walk traversal times are larger for networks with a higher degree
of
inhomogeneity in their shortcut distribution, and we discuss access time
distributions as functions of the initial and final node degrees. These
findings are relevant, in particular, when considering the optimization
of
existing information networks by the addition of a small number of fast
shortcut connections.
\end{abstract}

\maketitle

\section{Introduction}

Empirical observations performed recently on real networks as different
as
the Internet and the World Wide Web, ecological and food webs, power
grids
and electronic circuits, genome and metabolic reactions, collaboration
among
scientists and among Hollywood actors, and many others, have shown some
striking similarities in their structure and topology \cite{alb02,
dor03}.
In particular, the observations have revealed that the degree
distributions
of these networks are fat-tailed and typically close to a power-law, $%
P(k)\sim k^{-\gamma}$, where the exponent is usually in the range $%
2<\gamma<3 $ \cite{dor03,dor02a}.

These findings, exciting because of their near-universal occurrence,
have
motivated numerous investigations on the topology and geometrical
properties
of scale-free (SF) and other types of complex networks, as well as many
studies on dynamical and critical phenomena of statistical systems
defined
on complex network structures. Some of these efforts have been devoted
towards understanding the transport properties of discrete time and
continuous time random walks defined on different kinds of complex
networks %
\cite{nos87,sca01,pan01,lah01,alm03,noh04,yan05,par05,cos07}.

In particular, random walks have been studied on small- world networks
(SWNs), which are structures formed by superimposing a classical random
graph component (formed by randomly distributed long-range links or
shortcuts) on a regular lattice \cite{wat98,wat99}. Among the recent
contributions to this topic, Parris and Kenkre \cite{par05} introduced
the
important new feature of considering different jump rates for steps
across
regular lattice bonds ($F$) and across network shortcuts ($f$). This
type of
model is relevant to attempts to design, modify or optimize existing
information networks in order to reduce the mean access time, by
incorporating a small number of fast connections into an existing
network
already possessing a large number of other, perhaps slower, connections.

Ref.~\cite{par05} focused on the \textit{traversal time}, defined
as
the mean first-passage time for a random walker to reach the site
furthest
from its starting point, as measured along the 1D lattice (ring)
backbone,
averaged over different initial sites and network realizations. In this
earlier study, a collapse of traversal times was found, with interesting
universal behavior for $f/F\gg 1$ and low shortcut densities. By the
term
shortcut density $n_{sw}$ we mean the ratio of the number of shortcut
bonds
in the system to the total number of sites in the network, i.e.,
one-half
the average shortcut degree $\bar{k}$. In addition to the collapse, a
crossover was observed at a critical density of shortcuts
$n_{sw}=\bar{k}%
/2\approx 1$, which was attributed to percolation of the random graph
component of the network \cite{par05}.

The aim of the present work is to extend the scope of these previous
investigations by exploring random walks on hybrid network structures
formed
by superimposing SF networks onto regular lattices. We find that the
data
collapse and universal behavior previously observed with fast but low
shortcut densities on SWNs are also observed in this type of hybrid
structure, and that, consistent with the conclusions of Ref.
\cite{par05},
the crossover behavior associated with shortcut percolation is generally
present in networks satisfying the Molloy-Reed criterion
\cite{mol95,aie00};
such a transition is not seen in networks lacking a giant component.
Furthermore, we find that an inhomogeneous distribution of links of the
type
that occurs in SF networks leads to generally larger traversal times
when
compared to networks with links distributed homogeneously, a result that
bears resemblance to previous findings
\cite{lovasz,tetali,lopez,korniss} on
the effects of network heterogeneity on two point resistances and
synchronization efficiencies. Finally, we extend our study from the
traversal
times considered in Ref.~\cite{par05} to more general access times,
considered as functions of the initial and final node degrees, and
discuss
further the role played by hubs in this context.

The rest of the paper is laid out as follows. In Section II, we
introduce
the basic model, describe the algorithm that we use to generate the
hybrid
networks that form the subject of this investigation, and set out our
approach to studying the dynamics. In Section III, we present our
results
and a discussion. Section IV consists of concluding remarks.

\section{Model and Approach}

\begin{figure}[tp]
\begin{center}
\epsfxsize=3.4truein\epsfysize=2.5truein\epsffile{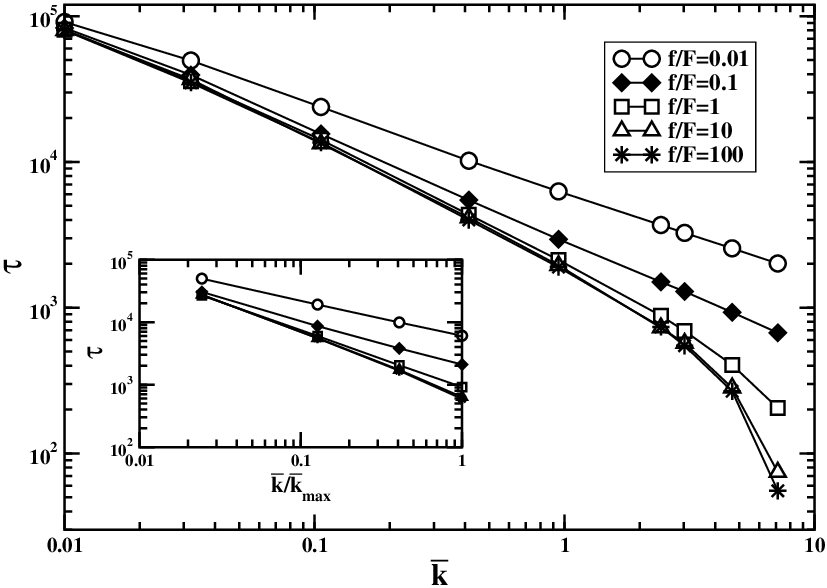}
\end{center}
\caption{Mean traversal time $\protect\tau $ as a function of the mean
shortcut degree $\bar{k}$ for different hopping rates ratios in the
range $10^{-2}\leq f/F\leq 10^{2}$. These results correspond to hybrid SF/RL
networks of size $N=1000$, where the SF component has exponent $\gamma =2$ 
and minimum degree $k_{0}=3$ (main plot) and $\gamma
=4$, $k_{0}=1$ (inset).}
\label{fig1}
\end{figure}

As stated in Section I, the aim of the present work is to explore the
dynamics
of random walks on hybrid network structures formed by superposing
shortcut
connections onto regular lattices. As in Ref.~\cite{par05}, the
random walks considered here occur in continuous time but, unlike in that
previous work, our interest here is in superposing \emph{scale-free}
structures. We thus provide below an explicit description first of how
we
build the networks and then of how we investigate the dynamics.

Each network of interest to the present study begins with a 1D lattice
of $N$
nodes (or sites) formed into a ring. Each node on this lattice is
connected
to each of its two immediate neighbors by a ``regular'' bond (or
connection). On top of this translationally invariant substrate we build
a
hybrid network by algorithmically connecting certain pairs of nodes with
``shortcut'' connections. With shortcuts drawn using a power-law degree
distribution, $P(k)\sim k^{-\gamma }$, a scale-free network of shortcuts
can
be superimposed on the underlying ordered ring, thus forming a hybrid
scale-free/regular-lattice (SF/RL) network. Besides the exponent $\gamma
$,
the degree distribution of a finite SF network is defined by the minimum
degree $k_{0}$ and the maximum (or cutoff) degree $k_{cut}$ of the nodes
participating in the network. Thus, in a normal ``stand alone'' scale
free
network, the number of sites with degree $k$ is 
\begin{equation} 
N_{k}^{sf}=
\left\{\begin{array}{cl}(k/k_{cut})^{-\gamma }
\ \ \ \ \text{for\ }k_{0}\leq k\leq k_{cut}\\ 0
\ \ \ \ \ \ \ \ \  \text{otherwise\ }\end{array}\right.\ ,
\label{dist}
\end{equation}   
the total number of nodes is
\begin{equation}
N^{sf}=\sum_{k=k_{0}}^{k_{cut}}N_{k}^{sf},
\end{equation}%
and the total degree is
\begin{equation}
K=\sum_{k=k_{0}}^{k_{cut}}kN_{k}^{sf}.
\end{equation}%

In embedding a SF network of \emph{shortcuts} into an already existing
regular lattice of $N$ nodes, we find that the SF network parameters are
restricted by the requirement that the number of nodes interconnected by 
shortcuts be less than or equal to the total number of nodes on the
lattice,
i.e., $N^{sf}\leq N$. For a given lattice size $N,$ this puts an upper
bound on the value of the average shortcut degree $\bar{k}=K/N$ that can be
realized.
For the range of parameters and systems sizes studied here, the maximum
average shortcut degree was bounded from above by values in the range
$1<\bar{k}_{max}<10$.

A finite SF network may be characterized either through the original
parameters $\gamma ,k_{0},$ and $k_{cut}$, or through the parameters
$\gamma,k_{0},$ and $\bar{k},$ replacing $k_{cut}$ by $\bar{k}$ through the
above relations. Our interest here being in comparing complex networks with
the same average shortcut degree $\bar{k}$, we employ the second parameter
triad ($\gamma ,k_{0},\bar{k}$) for network characterization.

In keeping with these ideas, to generate a particular realization of a
hybrid SF/RL network with a given set of parameters ($\gamma
,k_{0},\bar{k}$),
we randomly choose the lattice positions of the $N^{sf}$ nodes that will
belong to the SF shortcut subnetwork, and assign to each a given
shortcut degree $k$ following the power-law degree distribution (\ref{dist}). To
each of the $N_{k}^{sf}$ sites assigned shortcut degree $k$ we associate $k$
shortcut link ends, and then randomly connect pairwise the $K$ link
ends to establish the shortcut network (see e.g. \cite{her04}). 
This procedure, which is based on the so-called configuration model (see e.g. \cite{mol95,mol98,new03}) 
but includes an additional restriction on the maximum possible degree of the vertices, 
was shown to generate scale-free networks with no two- and three-vertex correlations \cite{bog04,cat05}. 

We now describe our approach for the analysis of random walks on hybrid
SF/RL networks generated as explained above. Our interest being in
continuous time random walks, as in Ref.~\cite{par05}, we could focus
on the Master equation, which contains information about the network
structure and the relevant hopping rates, numerically determine the
set of propagators, or Green's functions, and obtain from them the desired
transport quantities, following the method of Ref.~\cite{par05}.
Rather than pursue such a procedure, in the present paper we adopt for
computational convenience a simpler Monte Carlo approach. 

In the Monte Carlo calculations presented below, jump destinations of
the random walker are chosen at each time step from local transition
probabilities at the site occupied by the walker, by means of pseudo-random variables
(see e.g. \cite{bin02}). In particular, with transition rates $F$
associated with jumps to neighboring sites on the ordered ring, and transition
rates $f$ describing jumps between pairs of sites connected by a shortcut, a
walker located at a given site on the network with shortcut degree $k$ will
make a transition to one of its two neighbors on the ordered ring with
probability $p_{lat} = FT$, and will make a transition to one of the $k$ sites to which it is
connected by a shortcut with probability $p_{sh} = fT$, where $T$ is the duration 
of one Monte Carlo time step in the underlying continuous time random walk. 
The probability for the random walker to stay at its present position is thus $p_{stay} = 1-(2F+kf)T$. 
After performing $n_{MC}$ Monte Carlo steps, the corresponding 
continuous time elapsed is $\tau = Tn_{MC}$. 
The choice of $T$ is arbitrary to the extent of keeping all transition probabilities 
positive definite (i.e. $T\leq\left(2F+k_{cut}f\right)^{-1}$). 
We have explicitly verified that this method accurately reproduces the results
of the Master equation approach when applied to the SWN systems of Ref.~\cite{par05}.

\begin{figure}[t]
\begin{center}
\epsfxsize=3.4truein\epsfysize=2.5truein\epsffile{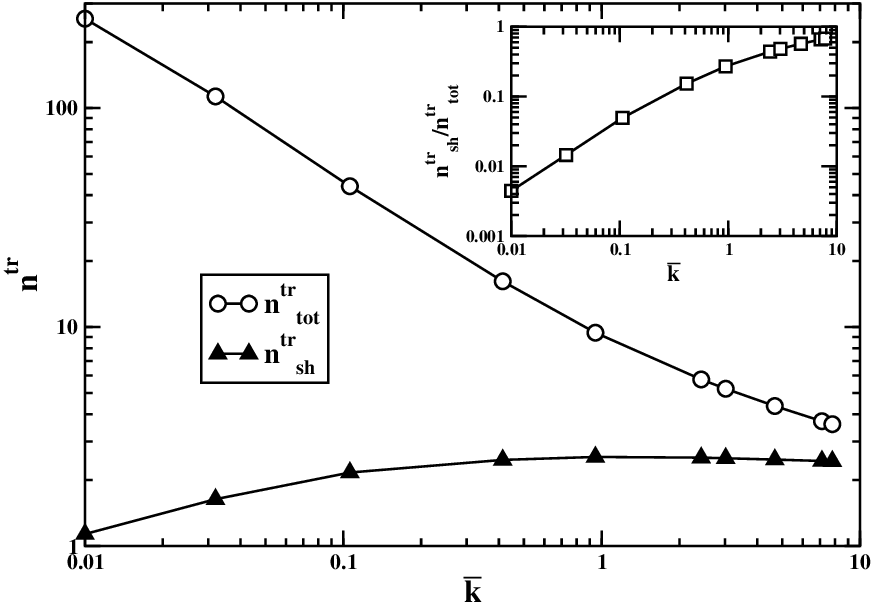}
\end{center}
\caption{Average shortest traversal path as a function of the mean
shortcut degree $\bar{k}$, where $n^{tr}_{tot}$ 
refers to the total number of steps, while $n^{tr}_{sh}$ is the number of 
steps taken across shortcut connections. Results correspond to hybrid SF/RL
networks of size $N=1000$, where the SF component has exponent
$\gamma =2$ and minimum degree $k_{0}=3$. 
The inset shows the ratio $n^{tr}_{sh}/n^{tr}_{tot}$ vs $\bar{k}$ for the same 
networks.} 
\end{figure}

As in Ref.~\cite{par05}, the focus on the present work is on the mean
time for a walker to traverse the system, and on more generally defined
{\it access} times, both of which are more straightforward to calculate using
the Monte Carlo procedure of the present paper than the Master equation.
Indeed, for each random walk trajectory, the traversal time $\tau $ is simply
the elapsed time between the moment the walk starts, and the moment it
arrives for the first time at the point $N/2$ sites away from which it started,
measured in either direction around the ring. More generally, we can
define the access time $\tau _{m,n}$ as the corresponding time for a walker
starting at site $m$ to arrive for the first time at site $n$. The mean
access time, for a given network configuration, is then the average of
this quantity over random walks starting from the same point on the same
network, and then over the ensemble of networks characterized by the same set of
network parameters. Except where explicitly noted, the numerical results
shown in the present paper were obtained for networks of size $N=1000$,
typically averaged over 100 different network configurations for each
set of network parameters, and over 1000 different random walk trajectories per
configuration.

\section{Results and Discussion}

Figure 1 shows the mean traversal time $\tau $ as a function of the mean
shortcut degree $\bar{k}$ for different hopping rates ratios in the
range $10^{-2}\leq f/F\leq 10^{2}$, as indicated. Here and throughout,
traversal times are measured in units of $F^{-1}$, which is the timescale for
jumps along regular lattice bonds. The results shown in the main plot
correspond to hybrid SF/RL networks in which the SF component has exponent $\gamma=2$
and minimum degree $k_{0}=3$, while the inset corresponds to $\gamma =4$
and $k_{0}=1$. Notice that the adoption of a particular SF degree
distribution, with given values of the parameters $\gamma $, $k_{0}$ and $k_{cut}$,
defines the total number of shortcuts, and hence the value of the mean
shortcut degree $\bar{k}=K/N$. This clearly differs from the SWN case, in
which the mean shortcut degree is a free parameter. In order to generate plots
of traversal times as functions of the shortcut density, we fix both
$\gamma $ and $k_{0}$, and consider different values of $k_{cut}$ under the
condition $N^{sf}\leq N$.

\begin{figure}[t]
\begin{center}
\epsfxsize=3.4truein\epsfysize=2.5truein\epsffile{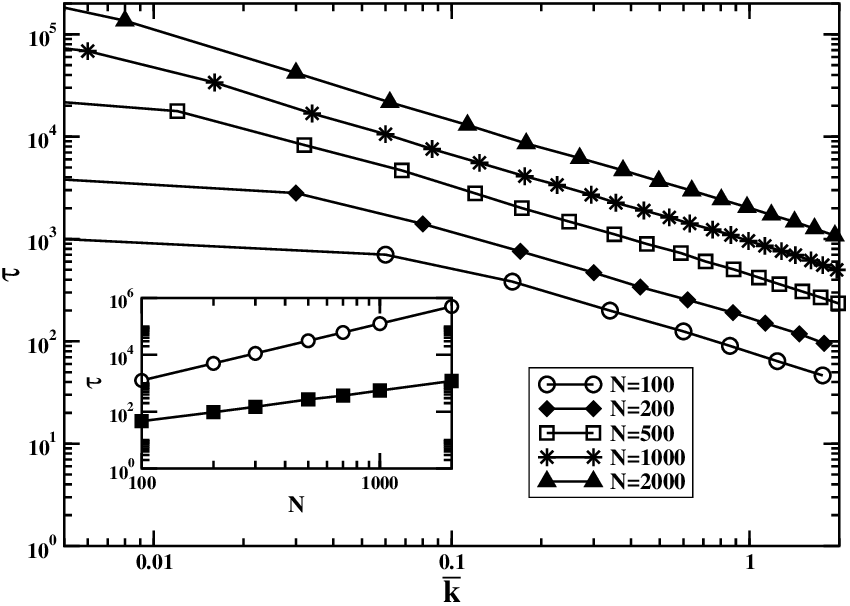}
\end{center}
\caption{Mean traversal time $\protect\tau $ as a function of the mean
shortcut degree $\bar{k}$ for networks of different size $N,$ as
indicated.
The inset shows the mean traversal time as a function of network size
$N$ for two systems, one with $\bar{k}=0$ (open circles), which scales as $N^{2}$ and
one with $\bar{k}=1.8$ (filled squares), which is very strongly connected, and for which
the traversal time scales as $N$. In both we have taken $f/F=1$.}
\end{figure}

Figure 1 shows that the mean traversal time decreases monotonically as
the mean shortcut degree increases, since a larger density of shortcuts
naturally contributes to shorten the time needed to random walk across
the network. The effect of decreasing the traversal times is only modest for
small values of the rates ratio $f/F$, but is increasingly large for
fast shortcuts ($f/F\gg 1$) and large shortcut densities close to the maximum
degree ($\bar{k}\simeq \bar{k}_{max}$).

In the limit $\bar{k}\ll 1$, the shortcut density is very low and the
traversal time tends to the diffusive limit $\tau_{diff}=
(1/2F)\times(N/2)^2 $. Even for $f/F\gg 1$, the fast shortcut
connections are sparse and do not lead to a substantial reduction of the traversal
time. This explains the data collapse onto a universal curve. However, when
the number of shortcuts is increased above a threshold close to
$\bar{k}\simeq\bar{k}_{max}$, which is related to the percolation of the shortcut
network component of the hybrid structure, the transport mode changes from being
dominated by diffusion on the lattice to being mainly due to propagation
along fast shortcut connections. As pointed out above, this behavior is
qualitatively similar to the mean traversal times observed in SWN and
other network structures \cite{par05}.

The condition for having a percolation threshold in a complex network is
that the degree distribution $P(k)$ satisfies
\begin{equation}
\sum_{k}k(k-2)P(k)>0\ ,  \label{molloy}
\end{equation}
which is known as the Molloy-Reed criterion \cite{mol95}. This equation
implies that the giant connected component is present for scale-free
networks with $k_{0}>1$, irrespective of $\gamma $. However, if
$k_{0}=1$, a
finite percolation threshold exists only if $\gamma <3.479$ \cite{aie00}.
The inset in Figure 1 shows traversal times corresponding to $\gamma =4$
and $k_{0}=1$, where the data are indeed observed to collapse onto a
universal curve for $f/F\gg 1$, even for $\bar{k}\approx \bar{k}_{max}$. Since the
degree distribution is very steep, however, in this case we are
restricted to $k_{cut}\leq 5$: only a few data points can be calculated.

\begin{figure}[tbp]
\epsfxsize=3.1truein\epsfysize=4.3truein\epsffile{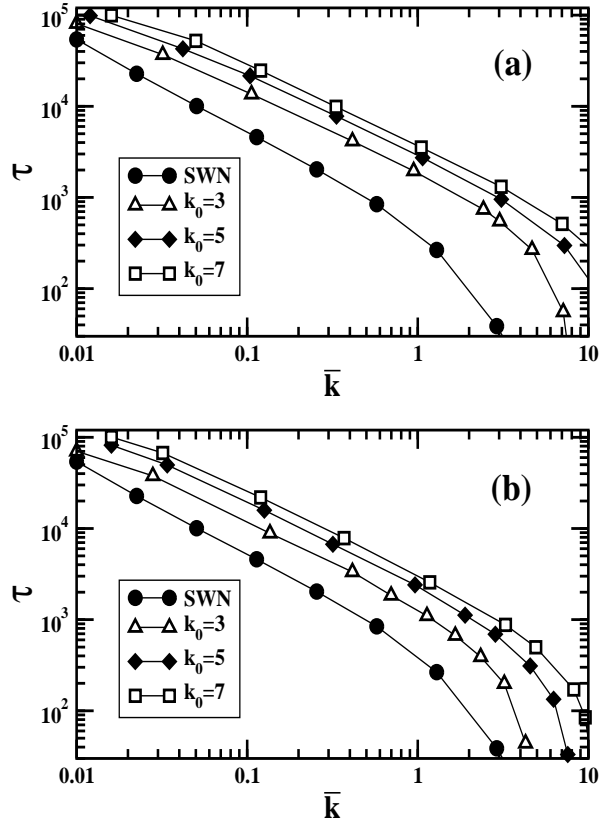}
\caption{Mean traversal time as a function of the mean shortcut degree,
as
obtained in the fast shortcut regime ($f/F=100$) for networks of size $%
N=1000 $. The traversal times for hybrid SF/RL networks with different
values of minimum degree ($k_{0}=3,5,$ and 7) are compared to
corresponding
SWN results. The values used for the exponent of the SF degree
distribution
are: (a) $\protect\gamma =2$ and (b) $\protect\gamma =3$.}
\label{FIG. 4}
\end{figure}

Figure 2 shows the average shortest traversal path as a function of the mean
shortcut degree $\bar{k}$, for hybrid SF/RL
networks of size $N=1000$, where the SF component has exponent
$\gamma =2$ and minimum degree $k_{0}=3$. We define $n^{tr}_{tot}$ as the total 
number of steps involved in the shortest path for traversing the system, averaged 
over different initial sites and different network configurations. 
Analogously, we can define $n^{tr}_{sh}$ ($n^{tr}_{lat}$) as the number of steps 
across shortcut connections (lattice bonds) corresponding to the same average 
shortest paths (such that $n^{tr}_{tot}=n^{tr}_{sh}+n^{tr}_{lat}$). 
The ratio $n^{tr}_{sh}/n^{tr}_{tot}$ is shown in the inset in Figure 2 
as a function of $\bar{k}$. This plot makes evident, particularly for large 
shortcut densities, the increasing importance of shortcut connections for 
traversing the system optimally.

Figure 3 shows how the mean traversal time $\tau $ as a function of $\bar{k}$ 
scales with the size of the network. We display in the
main figure $\tau$ vs $\bar{k}$ plots for various values of the network size
$N$. We display in the inset the mean traversal time as a function of
network size $N$ for two specific systems, one with $\bar{k}=0$ (open circles), which
scales as $N^{2}$ appropriate to the diffusive limit, and one with
$\bar{k}=1.8$ (filled squares), which is very strongly connected, and for which the traversal
time scales as $N$. In both we have taken $f/F=1$.

It is certainly of great interest to examine and possibly identify
correlations between structural network features and their relative
efficiency in signal transmission. With this aim, let us here compare
quantitatively the SWN case to hybrid networks with different SF
components.

Figure 4 shows a comparison of traversal times calculated for SWNs, which were
studied in Ref. \cite{par05}, with our new results
for hybrid SF/RL networks with different values of the minimum degree
($k_{0}=3,5,$ and 7) and two different exponent values: (a) $\gamma =2$
and (b) $\gamma =3$. In this figure, all networks have the same total number
of nodes, and are plotted as a function of the mean shortcut degree
$\bar{k}$, which for fixed $N$ is a direct measure of the total number of
shortcuts $N_{s}=N\bar{k}/2$ in the network. Thus, by comparing traversal
times of different networks of the same size at a given value of $\overline{k}$
we are able to address the question of how best to optimize the traversal
time with a fixed number of shortcuts. The results displayed in this figure
all correspond to $f/F=100$, which is the relevant limit in the problem of
network optimization through the addition of fast shortcuts.

The plots clearly reveal that, for a given value of the shortcut
density, the SWNs yield generally lower traversal times. Hence, we conclude that
inhomogeneously distributed shortcuts yield larger mean traversal times,
when compared to the case of networks with homogeneously distributed
shortcuts. Moreover, the traversal times for hybrid SF/RL networks show
a monotonic dependence with their corresponding minimum degree: the larger
$k_{0}$ (and thus, the more concentrated the shortcuts in a relatively
small number of highly connected nodes), the larger the mean traversal times.
These observations clearly indicate the key role played by the degree of
inhomogeneity of the shortcut distribution in the network's transport
properties.

Further insight into the role of shortcut distribution inhomogeneities
can be gained by studying mean \textit{access times} as functions of initial
and final node degrees. We can define the mean access time $\tau_{i,k}$ as
the
average first passage time for going from an arbitrary initial site with
degree $i$ to an arbitrary final site with degree $k$. In particular,
let us
consider access times $\tau_k^0\equiv\tau_{0,k}$ for initial sites
without
shortcut connections (i.e. sites that belong to the backbone regular
lattice, but do not participate in the SF component), and 
$\tau_k^{hub}\equiv\tau_{k_{cut},k}$ for the case in which the initial
site is the SF hub (i.e. the network's most connected site).

\begin{figure}[tbp]
\epsfxsize=3.4truein\epsfysize=2.5truein\epsffile{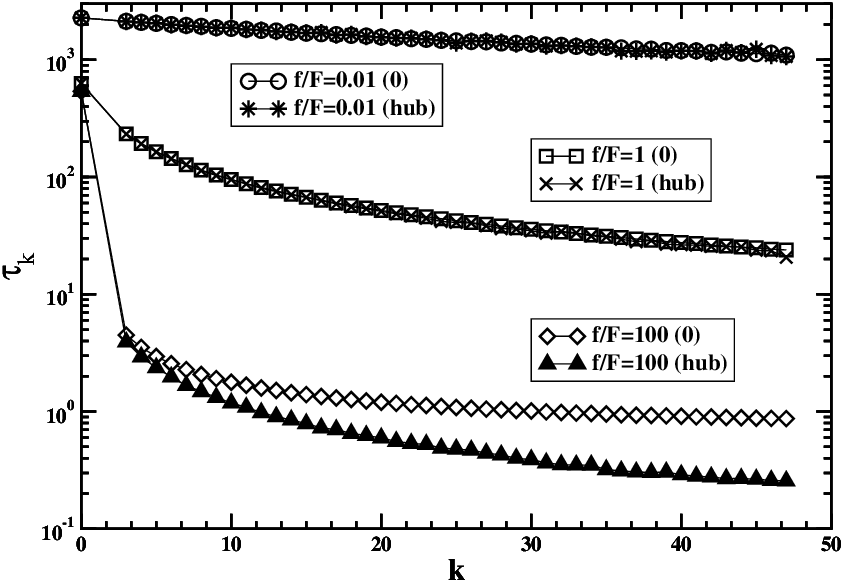}
\caption{Mean access times $\protect\tau_k^0$ and $\protect\tau_k^{hub}$
as
functions of the final node degrees, for hybrid SF/RL networks with
$\gamma=2$, $k_0=3$, $k_{cut}=49$ and different hopping rates ratios, as
indicated. See more details in the text.}
\label{FIG. 5}
\end{figure}

Figure 5 shows mean access times $\tau_k^0$ and $\tau_k^{hub}$ as
functions
of the final node degrees, for hybrid SF/RL networks with $\gamma=2$,
$k_0=3$%
, $k_{cut}=49$ and different hopping rates ratios in the range
$10^{-2}\leq
f/F\leq 10^2$. Notice that, for these network parameters, the resulting
SF
shortcut structure is quite dense ($N^{sf}/N=0.9$ and $\bar{k}=7.1$).
For $%
f/F\leq 1$, it is observed that $\tau_k^0\simeq\tau_k^{hub}$
irrespective of
the degree of the final node. Indeed, this is the expected result for
the
case of relatively slow shortcuts, where the connections of the complex
network component play a minor role. However, for $f/F\gg 1$, it is seen
that $\tau_k^0 > \tau_k^{hub}$ for all $k\geq k_0$, which can be
explained
from the fact that walkers starting from isolated nodes need to take
some
steps along slow lattice bonds before reaching sites belonging to the
fast
shortcut SF component. Arguments along the same lines also explain the
$k-$%
dependence of these plots, and particularly the large gaps observed in
the $%
f/F\gg 1$ case between final nodes with degrees $k=0$ and $k\geq k_0$.

Finally, let us consider the minimum, mean and maximum access times, and
compare their behavior for different hopping rates ratios. For random
walks
starting at sites without shortcut connections, the minimum, mean and
maximum access times are respectively defined as
$\tau_{min}^0=\mathrm{min}\
{\tau_k^0\}}$, $\bar{\tau}^0=(k_{cut}-k_0+1)^{-1}\sum_k\tau_k^0$and $%
\tau_{max}^0= \mathrm{max}\{\tau_k^0\}$, with analogous definitions
holding
for walks starting at the SF hub.

\begin{figure}[tbp]
\epsfxsize=3.4truein\epsfysize=2.5truein\epsffile{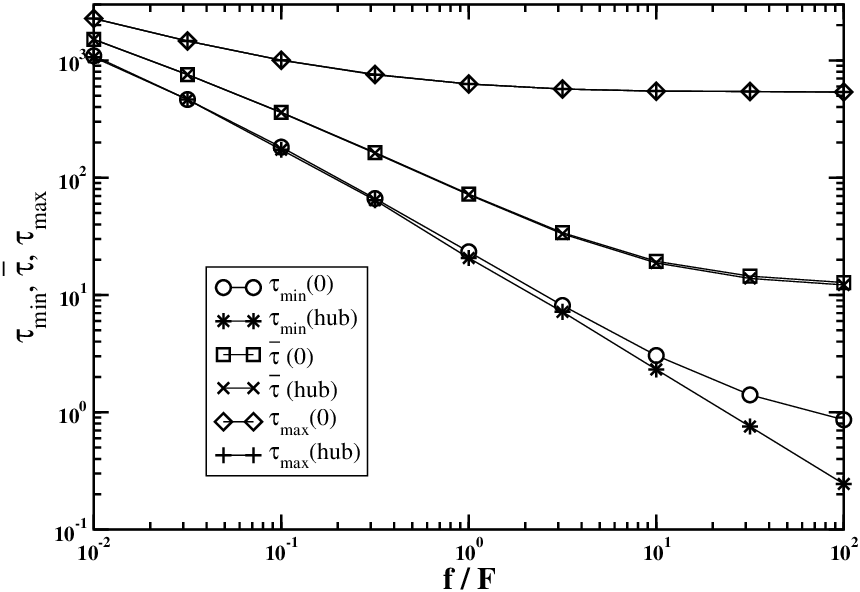}
\caption{Minimum, mean and maximum access times for random walks
starting at
sites without shortcut connections, as well as for walks starting at the
SF
hub. The hybrid SF/RL network parameters are $\protect\gamma=2$,
$k_0=3$, $%
k_{cut}=49$. See text for more details.}
\label{FIG. 6}
\end{figure}

Figure 6 shows these three access times as functions of the rates ratio
$f/F$%
, for hybrid SF/RL networks with parameters $\gamma=2 $, $k_0=3$, $%
k_{cut}=49 $. As expected from the previous discussion, little
difference is
seen in the case of slow shortcuts, since then the transport mode is
dominated by diffusion along lattice bonds. However, in the $f/F\gg 1$
regime, the minimum access times are found to differ substantially due
to
the existence of fast long-range shortcut connections among highly
connected
nodes.

\section{Concluding Remarks}

In summary, we have studied mean traversal and access times for random
walks
on hybrid scale-free/regular-lattice networks, and have considered
different
hopping rates for steps across lattice bonds and across network
shortcuts.
We have found two interesting results. First, in the limit of fast and
sparse shortcut connections, traversal times collapse onto a universal
curve. Second, a crossover behavior occurs related to the percolation
threshold of the scale-free network component for higher shortcut
densities.
We have discussed the occurrence of the transition in terms of the
Molloy-Reed criterion, which specifies the conditions for the existence
of
the giant connected component in the scale-free network structure.
Although
the qualitative behavior of traversal times is similar to observations
recently reported in Newman-Watts small-world networks \cite{par05}, the
quantitative comparison reveals that mean traversal times are larger for
networks with a higher degree of inhomogeneity in their link probability
distribution. Scale-free networks represent the paradigm for complex
random
structures possessing inhomogeneous probability distribution of links.
Our
purpose in comparing SF network results with small world network results
has
been to study the effect of inhomogeneities in the degree distribution
of
links. Finally, by considering access times as functions of the degree
of
the initial and final sites, we have stressed the key role played by
hubs in
reducing the networks minimum access times when fast shortcuts are
considered.

\section*{Acknowledgments}

We acknowledge useful discussions with Birk Reichenbach. This work was
supported in part by the NSF under grant no. INT-0336343. J. C. was also 
supported by the James S. McDonnell Foundation and the National 
Science Foundation ITR DMR-0426737 and CNS-0540348 within the DDDAS program.

\end{document}